\documentclass{osa-article}
\journal{oe}

\usepackage{siunitx}
\usepackage[utf8]{inputenc}
\usepackage{newunicodechar}
\newunicodechar{¹}{$^{1}$}
\newunicodechar{³}{$^{3}$}
\newunicodechar{⁴}{$^{4}$}
\newunicodechar{₃}{$_{3}$}
\newunicodechar{₄}{$_{4}$}

\articletype{Research Article}

\begin{document}

\title{Generating few-cycle pulses with integrated nonlinear photonics}

\author{David R. Carlson,\authormark{1,*} Phillips Hutchison,\authormark{1,2} Daniel~D.~Hickstein,\authormark{1} and Scott B. Papp\authormark{1,3}}

\address{\authormark{1}Time and Frequency Division, National Institute of Standards and Technology,\\325 Broadway, Boulder, CO 80305, USA\\
\authormark{2}Rhodes College, 2000 North Pkwy, Memphis, TN 38112, USA \\
\authormark{3}Department of Physics, University of Colorado, Boulder, CO 80305 USA}

\email{\authormark{*}david.carlson@nist.gov} 

\begin{abstract*}
Ultrashort laser pulses that last only a few optical cycles have been transformative tools for studying and manipulating light--matter interactions. Few-cycle pulses are typically produced from high-peak-power lasers, either directly from a laser oscillator, or through nonlinear effects in bulk or fiber materials. Now, an opportunity exists to explore the few-cycle regime with the emergence of fully integrated nonlinear photonics. Here, we experimentally and numerically demonstrate how lithographically patterned waveguides can be used to generate few-cycle laser pulses from an input seed pulse. Moreover, our work explores a design principle in which lithographically varying the group-velocity dispersion in a waveguide enables the creation of highly constant-intensity supercontinuum spectra across an octave of bandwidth. An integrated source of few-cycle pulses could broaden the range of applications for ultrafast light sources, including supporting new lab-on-a-chip systems in a scalable form factor.
\end{abstract*}

\section{Introduction}

Nonlinear optical interactions within a material rely on the ability to produce laser pulses with relatively high peak intensities. Achieving strong temporal and spatial confinement of the pulse is therefore important for optimizing the efficiency of a nonlinear process, especially for integrated-photonics applications where picojoule-energy pulses are used. To get the highest nonlinear response, simultaneous access to the few-cycle regime~\cite{kartner_few-cycle_2004,krausz_attosecond_2009} and wavelength-scale mode-field diameters is necessary. On-chip optical waveguides can readily provide the requisite spatial confinement due to their large refractive-index contrast, and the ability to easily control the spatial structure can be used to implement pulse compression~\cite{colman_temporal_2010,tan_nonlinear_2015}. However, group-velocity dispersion considerations make deterministically approaching the few-cycle regime a challenge experimentally.

In this work, we report the use of high-spatial-confinement nanophotonic waveguides made from stoichiometric silicon nitride (Si$_3$N$_4$, henceforth SiN) to access the few-cycle regime via the soliton self-compression effect. Moreover, we demonstrate how the nearly maximal spatial and temporal compression in the waveguide can be used for generating very constant, flat, and broadband spectra across octave bandwidths. The modest picojoule-level pulse energy requirements and the simple chip-integrated design may help make ultrashort laser pulses with peak intensities in the TW/cm$^2$ range available to new users and applications.

Soliton-effect pulse compression can occur for pulses propagating through a nonlinear medium with simultaneous self-phase modulation and anomalous group-velocity dispersion~\cite{agrawal_nonlinear_2006}. This approach to reducing the duration of a pulse is notable because it does not require any external dispersion compensation and can thus be simply implemented~\cite{foster_soliton-effect_2005, tognetti_sub-two-cycle_2007, kibler_optimised_2007}. For a given medium length, these compressors require a launched pulse with the appropriate characteristics (e.g. energy, duration, chirp) to achieve the maximal degree of temporal compression at the output of the medium~\cite{voronin_soliton-number_2008}. In practice, the necessary control can often be achieved by varying just a single parameter, such as the total pulse energy. 

\begin{figure*}
\centering
\includegraphics[width=0.84\textwidth]{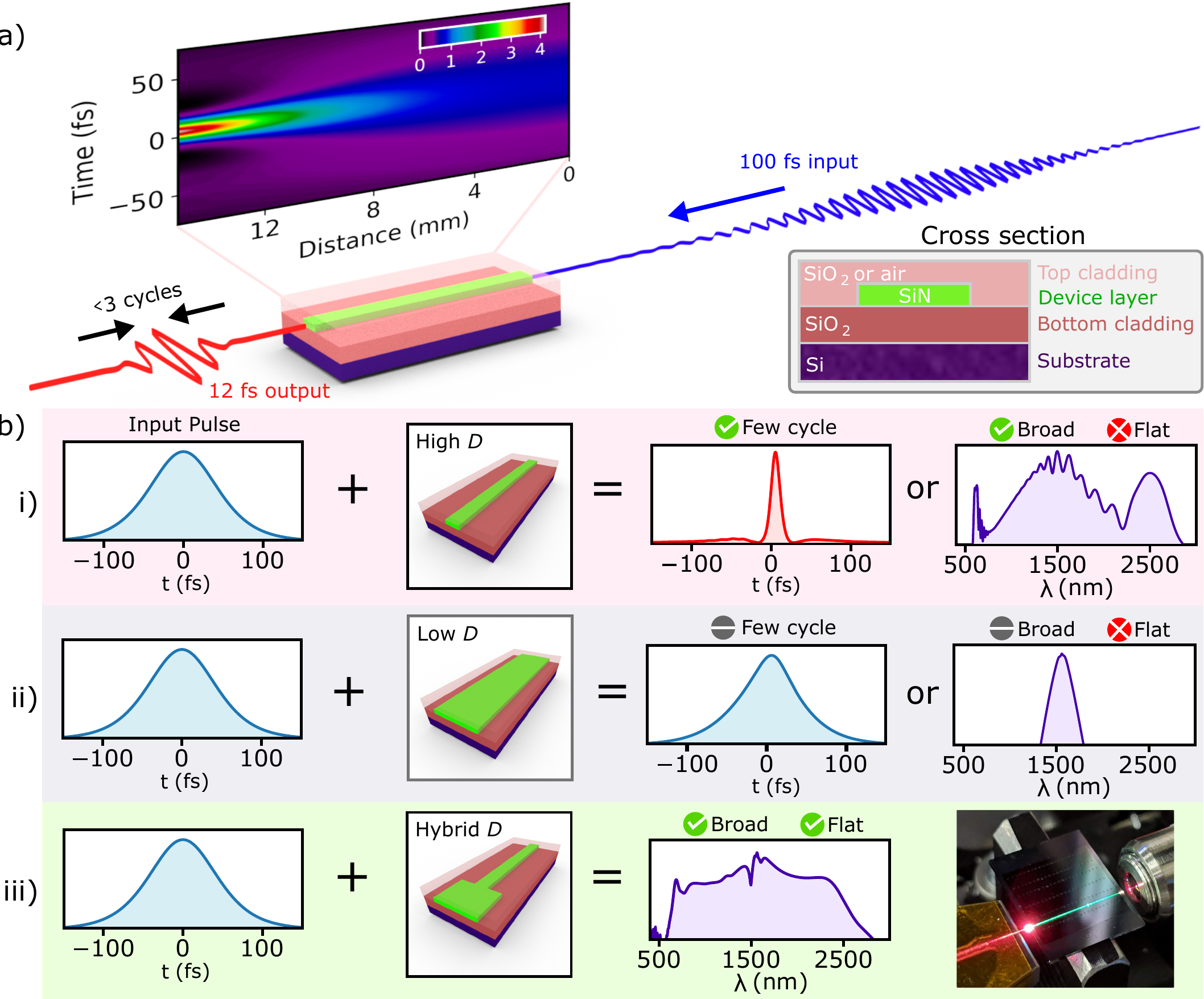}
\caption{\textbf{Few-cycle pulses and integrated photonics.} a) A nonlinear waveguide with anomalous dispersion and sufficient length can produce few-cycle pulses through soliton-effect compression. Right inset shows the material cross section. b) Propagation of a 100~fs input pulse is simulated in waveguides with different dispersion parameters $D$, which we control through the waveguide width. i) Narrow waveguides with high anomalous dispersion can be used to produce clean few-cycle pulses, but not output spectra that are simultaneously flat and broadband.  ii) If a wider waveguide, with low anomalous dispersion is used, the pulse does not undergo significant pulse compression or spectral broadening.  iii) Hybrid waveguides consisting of an initial segment of high dispersion for pulse compression followed immediately by a wider waveguide for spectral broadening can lead to highly flat and broadband output spectra. Far right panel shows a photograph of a two-section waveguide when supercontinuum generation occurs at the geometric boundary between segments.}
\label{fig1}
\end{figure*}

This concept, implemented here in a nanophotonic SiN waveguide, is shown in Fig.~\ref{fig1}a for a 100~fs full-width at half-maximum (FWHM) sech$^2$ input pulse at a center wavelength of 1550~nm. Under these conditions, the temporal and spectral characteristics of the output waveform can be dramatically influenced by the choice of waveguide geometry, as illustrated in Fig.~\ref{fig1}b. For example, a narrow waveguide with anomalous dispersion ($D\sim60$~ps/nm$\cdot$km) can cleanly compress the input pulse to 12~fs at 30~pJ or, at a higher energy of 70~pJ, can undergo soliton fission in the waveguide and produce a broad supercontinuum spectrum~\cite{halir_ultrabroadband_2012, epping_-chip_2015, johnson_octave-spanning_2015, carlson_self-referenced_2017, porcel_two-octave_2017}. While large dispersion values at the pump wavelength are desired for efficient pulse compression and broad bandwidths (i), they are generally not well suited for producing smooth and flat output spectra.  If a wider-width waveguide of the same length is used instead, the looser confinement and lower dispersion are insufficient to create either significant temporal compression or spectral broadening (ii). However, if the same wide waveguide is seeded with the few-cycle output of the narrow waveguide, a unique broadband and smooth spectrum can be achieved that is otherwise unattainable from a uniform waveguide geometry (iii).

\section{On-chip few-cycle pulse generation}

\begin{figure}
\includegraphics[width=0.7\linewidth]{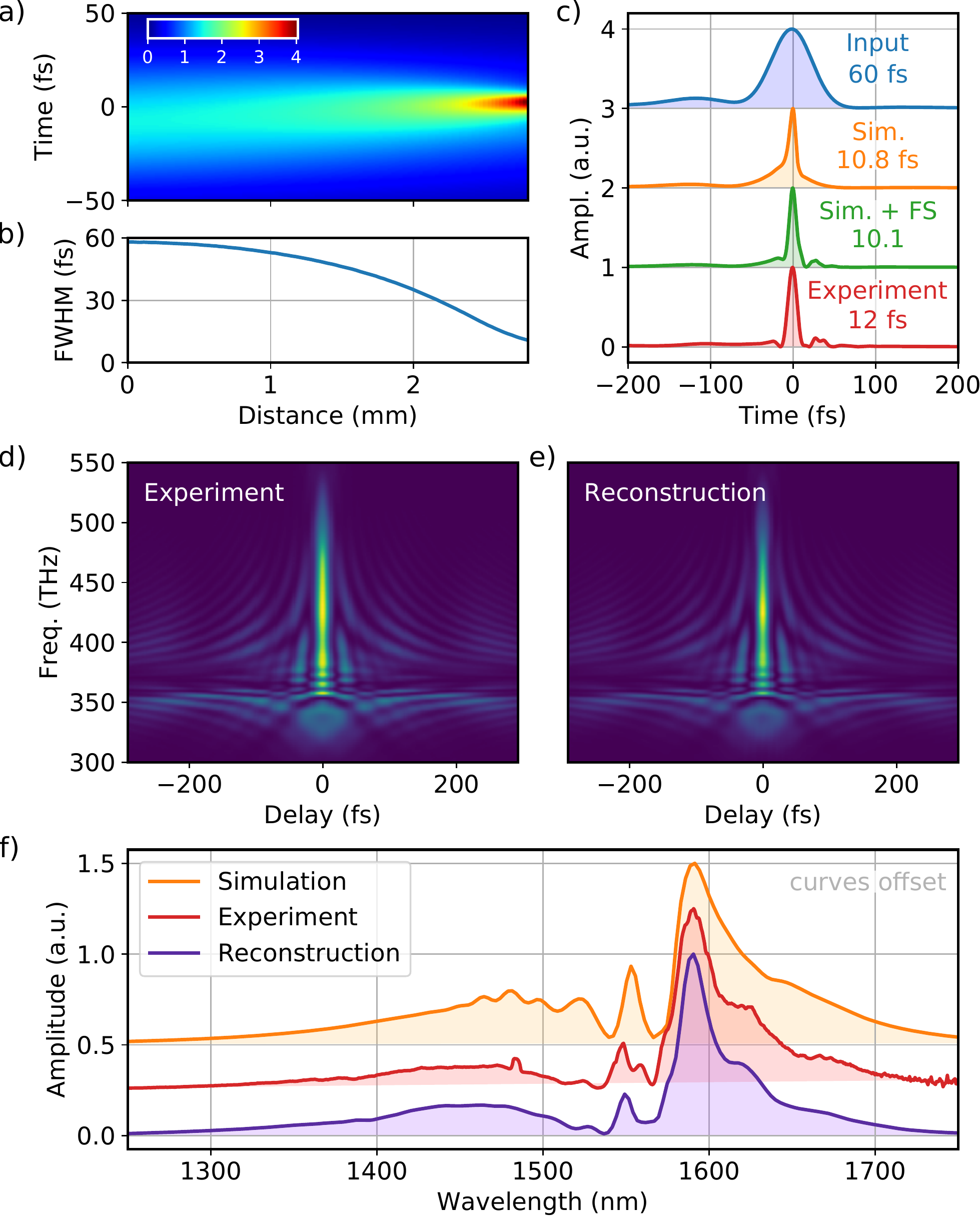}
\centering
\caption{\textbf{Few-cycle pulse generation in nanophotonic waveguides.} a) Simulated temporal profile and b) FWHM duration as a function of propagation distance in a straight waveguide with a width of 2.0~$\mu$m. c) Pulse profile for the input, experimental, simulated, and optimized pulse obtained with external propagation through 2.5~mm of fused silica. d) Experimental and e) reconstructed FROG traces recorded at the waveguide output. f) Spectra corresponding to the same pulses shown in panel c).}
\label{fig:fewcycle}
\end{figure}

To support our experimental measurements and gain insight into the underlying nonlinear dynamics, simulations of the nonlinear pulse propagation were performed using a split-step Fourier-method solver of the generalized nonlinear Schr\"odinger equation~\cite{ycas_pynlo:_2015}. The model includes a correction for wavelength-dependent mode area~\cite{kibler_supercontinuum_2005}, while the waveguide mode and dispersion characteristics were computed with the \texttt{wgmodes} solver~\cite{fallahkhair_vector_2008} using refractive-index data for SiN and SiO$_2$ from Refs.~\cite{luke_broadband_2015} and \cite{malitson_interspecimen_1965}, respectively. The fundamental quasi-transverse-electric mode of the waveguide is used in all cases.

Experimentally, we use a 1560~nm mode-locked fiber laser with a repetition rate of 100~MHz to seed the SiN waveguides~\cite{sinclair_invited_2015}. By varying the amplifier pump power, the transform-limited output pulse duration can be tuned between 50~fs and 300~fs. The waveguides themselves are fabricated commercially by Ligentec using low-pressure chemical vapor deposition (LPCVD) SiN, have a fully oxide-clad geometry, and include 200~nm inverse-taper structures at the chip facets for improved input and output coupling~\cite{almeida_nanotaper_2003}. The pulses collected from the waveguide are collimated with a reflective silver-coated parabolic mirror and directed to a homemade frequency-resolved optical gating (FROG) apparatus~\cite{trebino_measuring_1997}. The retrieved pulse is then characterized and compared to the model results, as shown in Fig.~\ref{fig:fewcycle} for a waveguide of width 2000~nm, thickness of 770~nm, and length of 2.8~mm.  When a 60~fs FWHM input pulse with an energy of 220 pJ is coupled to the waveguide, a pulse as short as 12~fs can be realized experimentally, in good quantitative agreement with the simulated temporal profile (Fig.~\ref{fig:fewcycle}c) and spectrum (Fig.~\ref{fig:fewcycle}f). However, in our implementation we note the inverse taper structure at the waveguide output can add significant group-delay dispersion to the pulse (GDD~$\sim$150~fs$^2$), despite having a total length of only \SI{300}{\micro\meter}. If the compressed pulse is to be used off-chip, then this effect can be readily compensated by introducing an appropriate amount of fused silica glass into the beam path ($\sim$10~mm for FROG trace in Fig.~\ref{fig:fewcycle}d).

The quality of pulse compression can be quantified by the ratio $Q$ of energy contained in the central pulse peak to the total energy in a sech$^2$ pulse with the same FWHM and amplitude, and is dependent on several factors including the waveguide dispersion and input pulse parameters. For example, the 60 fs pulses produced by our fiber laser have $Q$=0.76 and after compression in the waveguide (reconstruction in Fig.~\ref{fig:fewcycle}c) achieve a quality factor of $Q$=0.40. Under some conditions, introducing a small amount of anomalous dispersion can apparently improve the pulse quality without significantly affecting the FWHM duration. The simulated pulse in Fig.~\ref{fig:fewcycle}c, for instance, can see an increase in quality factor to 0.51 a 2.5 mm thickness window of fused silica is added to the off-chip beam path. Nevertheless, our calculations show that if a pure 60 fs, 30 pJ, sech$^2$ pulse instead seeds a 1~cm-long waveguide with the same transverse dimensions, the achievable output $Q$ value can reach 0.8 without introducing any additional material.

\section{Ultraflat supercontinuum with two-section waveguides}
\begin{figure}
\includegraphics[width=0.7\linewidth]{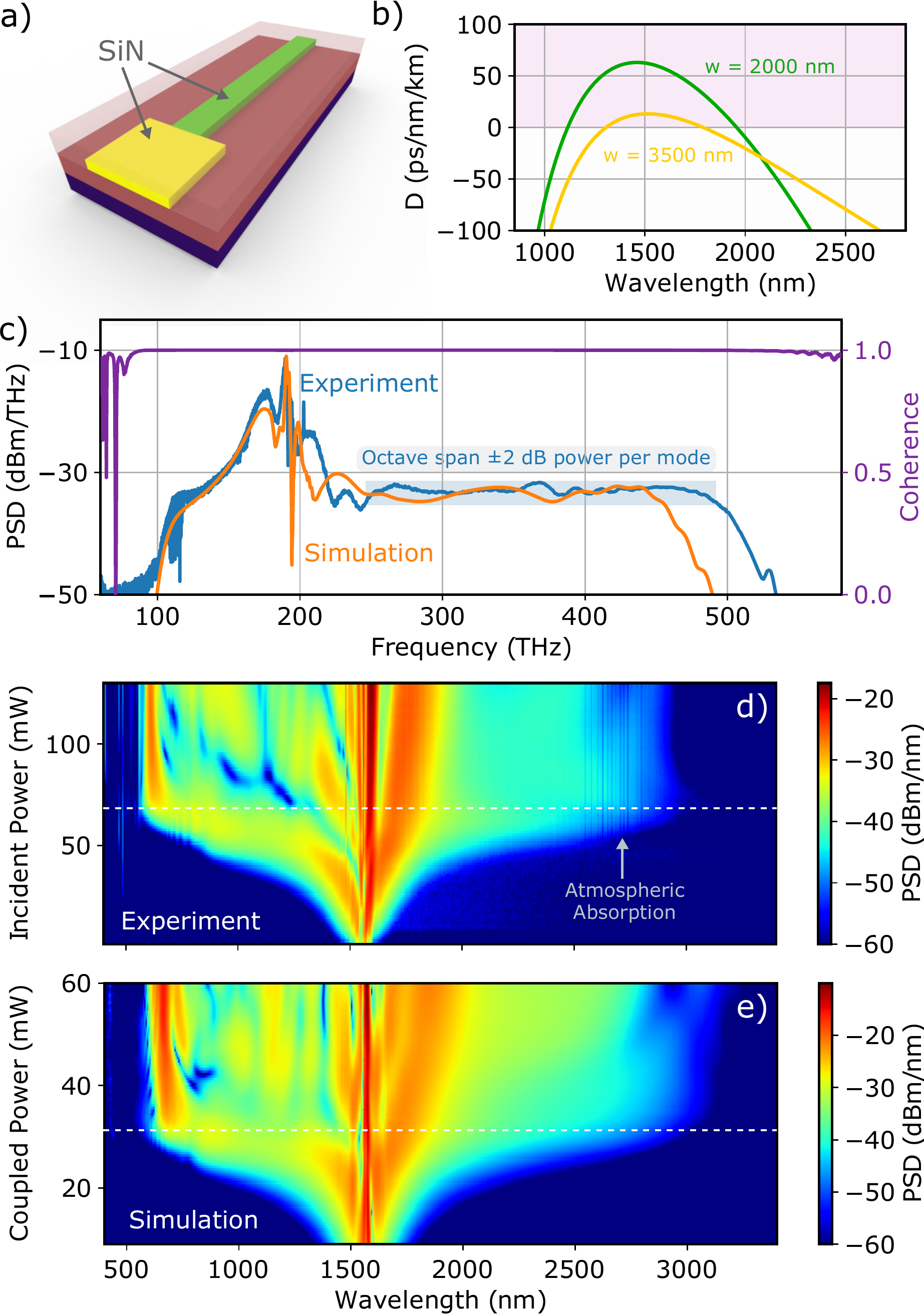}
\centering
\caption{\textbf{Ultaflat supercontinuum with width-changing waveguides.} a) Geometry layout and b) calculated dispersion profile for each waveguide section. c) The experimental output spectrum (blue) collected with a multi-mode fluoride fiber containing an octave-spanning spectral band with nearly constant power per mode. The simulated waveguide output (orange) shows that high spectral coherence (purple) is maintained across the full comb bandwidth. d) Experimental and e) simulated power scan showing the output spectrum vs incident pulse energy. The dashed white lines correspond to the curves shown in c).}
\label{fig:widthchange}
\end{figure}

The possibility of achieving maximal spatial and temporal confinement of a pulse in a waveguide structure may open new design possibilities for photonic devices. For example, here we show how seeding a low-anomalous-dispersion waveguide segment with the few-cycle output from an immediately preceding waveguide segment can produce very flat and broadband supercontinuum, without using passive~\cite{carlson_ultrafast_2018} or active~\cite{probst_spectral_2013} external flattening.  Fig.~\ref{fig:widthchange}a shows an implementation of this configuration, using an abrupt change to the waveguide width. As before, compression occurs during the first 2.8~mm of propagation in the 2000~nm-wide waveguide but undergoes soliton fission immediately upon entering the 3500~nm-wide segment for the final 0.5~mm of propagation.  The coherent output spectrum, shown in Fig.~\ref{fig:widthchange}c, yields a frequency comb containing an octave-spanning spectral band with a power-per-mode of $\sim$100~pW and a variation of only $\pm$2~dB. If a higher power per mode is desired, high-repetition-rate sources based on electro-optics~\cite{carlson_ultrafast_2018} or microresonators~\cite{lamb_optical-frequency_2018} could instead be used to drive the supercontinuum.

\begin{figure}
\includegraphics[width=0.7\linewidth]{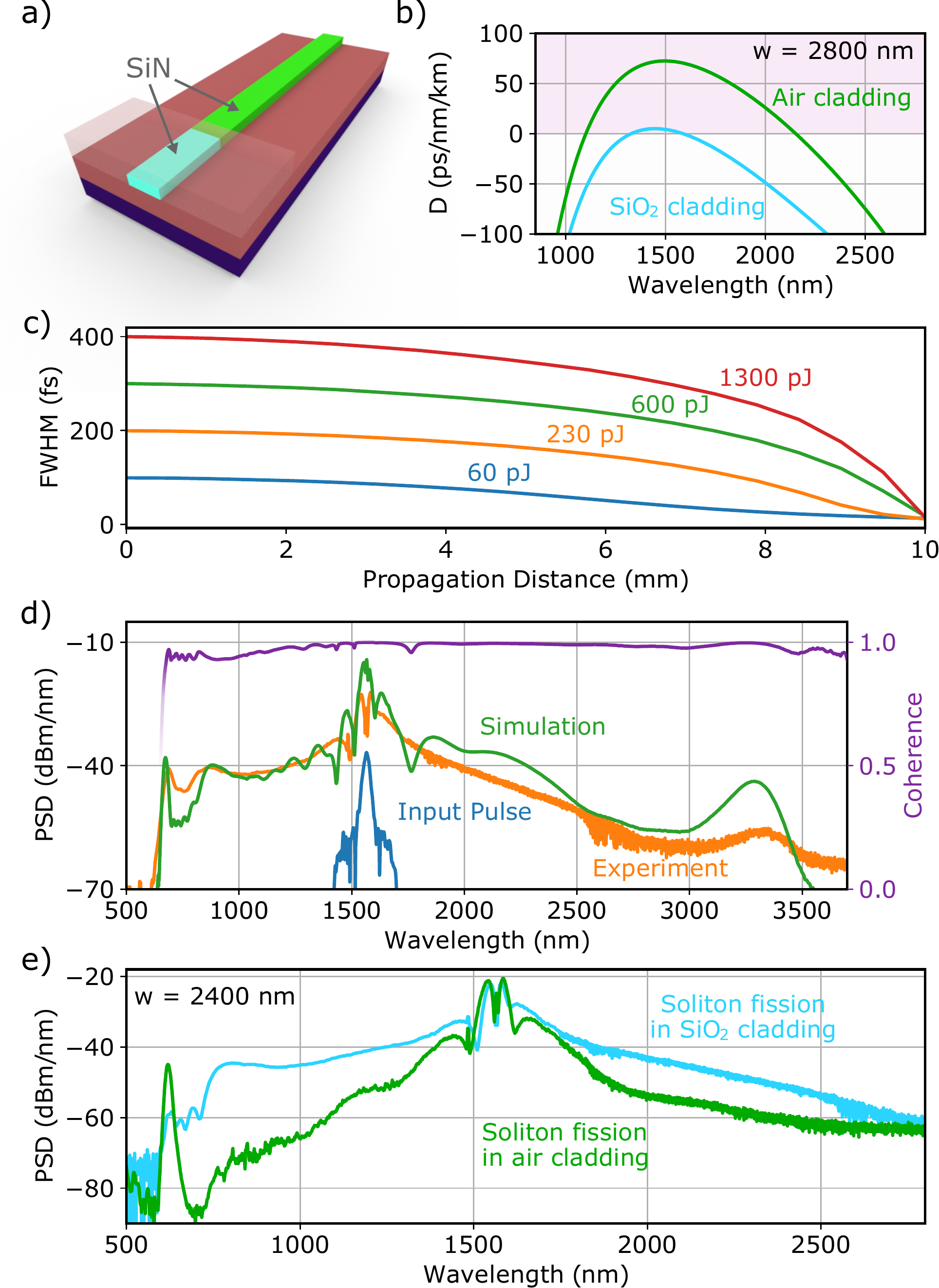}
\centering
\caption{\textbf{Pulse compression and supercontinuum with cladding-changing waveguides.} a) Geometry layout and b) calculated dispersion profile for each waveguide section. c) Simulated pulse duration in 1-cm long air-clad waveguides as function of propagation distance for input pulses ranging from 100 to 400~fs. d) Experimental and simulated output spectra when a 250~fs input pulse achieves maximal compression at the interface between cladding materials. Despite the long pulse, a high degree of coherence is maintained across the full spectral bandwidth. e) Experimental output spectra for a 2400~nm wide waveguide obtained by adjusting the input power by 6~\% to allow the pulse to fully undergo soliton fission in the oxide-clad region (blue) and the air-clad region (green).}
\label{fig:claddingchange}
\end{figure}

An alternative approach to controlling the waveguide dispersion is to change the top cladding layer of the waveguide.  In Fig.~\ref{fig:claddingchange}, we show how devices with an ``air-cladding'' geometry can be used in combination with a symmetric SiO$_2$ cladding region near the edge of the chip to again achieve two different dispersion profiles (Fig.~\ref{fig:claddingchange}b) in the same waveguide. This variant has the advantage of accessing the dispersion of air-clad devices while still maintaining high coupling efficiency due to the symmetric mode profile and larger spot size at the chip facets~\cite{carlson_self-referenced_2017}.

Supercontinuum generation with two-section waveguides consisting of pulse compression in a segment of high-anomalous dispersion followed by soliton fission in a lower-dispersion segment has been proposed as a method for coherent broadening of pulses with durations as large as 1~ps~\cite{johnson_coherent_2017}. To investigate if our devices can be used to compress longer-duration input pulses, the length of the 700~nm-thickness air-clad region in these devices is increased to 1~cm. Fig.~\ref{fig:claddingchange}c shows the simulated FWHM pulse duration evolution for input pulses ranging from 100 to 400~fs as a function of propagation distance.  Provided the pulse energy is adjusted appropriately, the same waveguide can ultimately achieve compression to the few-cycle regime in each case. For a 250~fs input pulse, Fig.~\ref{fig:claddingchange}d shows both experimental and simulated spectra for a cladding-change waveguide of width 2800~nm when maximum compression is achieved at the interface between cladding layers.

However, a common problem for supercontinuum with pulses $>$200~fs is maintaining the coherence of the broadened spectrum. Here we estimate the spectral coherence by simulating input-pulse shot noise through the addition of a ``one-photon-per-mode'' seed field with randomized phase to each frequency discretization bin~\cite{dudley_supercontinuum_2006, dudley_numerical_2002}. The modulus of the complex degree of coherence $|g_{12}|$ for 1,000 independent pulse pairs is then computed from the phase and amplitude variations of independent fields $E_1$ and $E_2$ via  
\begin{equation*}
	g_{12}(\omega) = \frac{\langle E_1^*(\omega) E_2(\omega) \rangle}{\sqrt{\langle |E_1|^2 \rangle\langle |E_2|^2 \rangle}},
\end{equation*}
where the angle brackets denote ensemble averages over all pulses.

Fig.~\ref{fig:claddingchange}d shows that for our 250~fs input pulse, $|g_{12}|$ remains above 0.9 over the full bandwidth of the output spectrum.  However, at this level there is already an obvious reduction in coherence compared to the 67~fs pulse used in Fig.~\ref{fig:widthchange} for the width-changing waveguide. While longer pulses approaching 1~ps can still produce a broad smooth continuum and compress to few-cycle durations, our results suggest that like most soliton-fission-based techniques, the coherence of the full output spectrum is not preserved.  Instead, a multi-stage scheme using normal-dispersion broadening before coupling into the waveguide would likely offer better performance~\cite{carlson_ultrafast_2018,lamb_optical-frequency_2018}. More generally, schemes based on nonlinear spectral broadening in the normal-dispersion regime~\cite{hooper_coherent_2011} that take advantage of ultralow-loss waveguides will be important to support emerging heterogeneously integrated, gigahertz-rate modelocked lasers~\cite{davenport_integrated_2018}.

An additional and related consequence of using longer-duration input pulses and a longer waveguide segment for pulse compression is that the spectrum exhibits a high sensitivity to input power fluctuations. For example, Fig.~\ref{fig:claddingchange}e shows two different experimental spectra from the cladding-change waveguide when the input power is varied by 6~\%.  At the higher power, soliton fission occurs completely within the air-clad region (green curve), producing the strong dispersive wave near 600~nm.  However, at the reduced power level, soliton fission occurs in the oxide-clad region (blue curve), producing a much flatter, albeit slightly narrower, spectrum with a maximum power difference of up to 30~dB near a wavelength of 800~nm.

Provided the input pulse to the waveguide can be well controlled, our results indicate that the design versatility inherent to integrated photonics allows access to the few-cycle-pulse domain in an on-chip platform.  Moreover, the efficient concentration of pulse energy in both space and time in these devices enables new possibilities for even relatively mature fields such as supercontinuum generation. Achieving high peak intensities with low pulse energies could also benefit high-speed opto-electronic switching or more extreme wavelength conversion techniques such as intrapulse difference-frequency generation to infrared wavelengths or solid-state high-harmonic generation to the extreme ultraviolet. Finally, the techniques demonstrated here should be widely applicable to other nonlinear photonic materials besides SiN, in support of the ever-diversifying ultrafast photonics landscape.

\section*{Acknowledgments}
We thank Kartik Srinivasan and Daron Westly for help with device fabrication, Scott Diddams for helpful insights, and Nima Nader and Abijith Kowligy for comments on the manuscript. This research is supported by the Air Force Office of Scientific Research (AFOSR) under award number FA9550-16-1-0016, the Defense Advanced Research Projects Agency (DARPA) DODOS program, the National Institute of Standards and Technology (NIST), and the National Research Council (NRC). Certain commercial equipment, instruments, or materials are identified here in order to specify the experimental procedure adequately. Such identification is not intended to imply recommendation or endorsement by the National Institute of Standards and Technology, nor is it intended to imply that the materials or equipment identified are necessarily the best available for the purpose. DDH is currently employed by KMLabs Inc., a company that manufactures femtosecond lasers. DRC is a co-founder of Octave Photonics, a company specializing in nonlinear integrated photonics. This work is a contribution of the U.S. government and is not subject to copyright. 

\bibliography{references}

\end{document}